# Direct Observation of Unusual Interfacial Dzyaloshinskii-Moriya Interaction in Graphene/NiFe/Ta Heterostructure

Avinash Kumar Chaurasiya[1,§], Akash Kumar[2,§], Rahul Gupta[2], Sujeet Chaudhary[2], Pranaba Kishor Muduli[2], and Anjan Barman[1,*]

[1]*Department of Condensed Matter Physics and Material Sciences, S. N. Bose National Centre for Basic Sciences, Block JD, Sec. III, Salt Lake, Kolkata 700106, India*

[2]*Thin Film Laboratory, Department of Physics, Indian Institute of Technology Delhi, Hauz Khas, New Delhi 110016, India*

E-mail: *abarman@bose.res.in
[§]Authors with equal contribution

## Abstract

Graphene/ferromagnet interface promises a plethora of new science and technology. The interfacial Dzyaloshinskii Moriya interaction (iDMI) is essential for stabilizing chiral spin textures, which are important for future spintronic devices. Here, we report direct observation of iDMI in graphene/$Ni_{80}Fe_{20}$/Ta heterostructure from non-reciprocity in spin-wave dispersion using Brillouin light scattering (BLS) technique. Linear scaling of iDMI with the inverse of $Ni_{80}Fe_{20}$ thicknesses suggests primarily interfacial origin of iDMI. Both iDMI and spin-mixing conductance increase with the increase in defect density of graphene obtained by varying argon pressure during sputter deposition of $Ni_{80}Fe_{20}$. This suggests that the observed iDMI originates from defect-induced extrinsic spin-orbit coupling at the interface. The direct observation of iDMI at graphene/ferromagnet interface without perpendicular magnetic anisotropy opens new route in designing thin film heterostructures based on 2-D materials for controlling chiral spin structure such as skyrmions and bubbles, and magnetic domain-wall-based storage and memory devices.

One of the key motivations of modern spintronics research is to achieve low power consumption, faster information processing and higher storage density in thin-film based magnetic memory devices. Ferromagnetic (FM) thin films adjacent to nonmagnetic thin layers give rise to a range of important phenomena, relevant for the emergent field of spin-orbitronics. These include perpendicular magnetic anisotropy (PMA) [1], spin pumping [2], spin torque [3], spin Hall effect [4], Rashba effect [5, 6], interfacial Dzyaloshinskii-Moriya interaction (iDMI) [7] and chiral damping [8]. While most of these phenomena are associated with high spin-orbit interaction, and heavy metal (HM) layers are natural choices for the nonmagnetic layer, emerging materials like two-dimensional materials and topological insulators may also play important roles in engineering the interface magnetism. The iDMI, is an anti-symmetric exchange interaction which originates due to the broken inversion symmetry at the HM/FM interface, where HM possess high spin-orbit coupling [7]. It favors canted spin configurations which gives rise to various magnetization structures at the nanoscale such as Skyrmions [9-11] and chiral helices [12]. In addition, it can improve domain wall velocities by suppressing Walker breakdown in magnetic racetrack memory devices and lead to non-reciprocal spin-wave propagation leading towards applications in the high-speed spin-wave logic device [13]. Recently, the direct observation of iDMI has been evidenced mainly in HM/FM/oxide heterostructures [14-25].

Graphene and other 2-D materials such as $MoS_2$ have shown promises in spintronics [26]. The fascinating properties of graphene such as massless Dirac Fermions in the linear dispersion of its electronic structure, well-defined monolayer formation, and long spin-diffusion length have created burgeoning interest in the scientific community for its applications in spintronics research [26-28]. The use of graphene in magnetic sandwich structures which aims to search for new materials have been described theoretically as well as experimentally in the context of magnetic tunnel junctions [29-34]. In recent years, it has been shown that graphene plays a key role in Rashba effect [5, 6], enhancement of spin injection efficiency [35, 36] and quantum spin Hall effect [37]. Moreover, there is sufficient experimental evidence that the formation of graphene-metal contacts significantly modifies the electronic and/or magnetic properties of the interface and investigations related to charge transfer from metal to graphene have also been reported [38-40]. In addition, the weak spin-orbit coupling (SOC) in $sp^2$ carbon also suggested that electron spin should be carried nearly unaffected over unprecedented distances, making feasible, practical applications of lateral spintronics [41,42]. The in-plane $sp^2$ bonding is mainly responsible for the structural stability and mechanical strength of graphene, whereas out-of-plane p$\pi$ states control its transport and interfacial properties. Pioneering works initially resounded such high expectations [43], while more recent experimental results have confirmed the potential of graphene for transporting spin signal at room temperature over tens of micrometers, which makes it a better candidate for technological realization [36, 44, 45].

There is a recent report on the giant enhancement of PMA in Co-graphene heterostructures despite graphene having a small spin-orbit interaction, indicating the unusual nature of the graphene-FM interface [46]. A low-damage high-throughput grazing-angle sputter deposition on graphene has been successfully demonstrated [47]. However, the promotion of graphene-FM heterostructures for applications of spin-orbitronics has not been intensely started yet. Recently, Yang *et al.* have reported the observation of significant DMI at the graphene-FM interface by employing both first principle calculation and magnetic imaging experiments [48]. Most recently, Ajejas et al have reported the existence of sizeable DMI at gr/Co interface from the Kerr experiments [49]. The physical origin of the observed DMI in both Ref. 48 and 49 was attributed to the conduction electron mediated Rashba effect originated at the graphene-FM interface. Yang *et al.* used Co films grown by molecular beam epitaxy on chemical vapor deposited graphene on Ru (0001) substrate to demonstrate the DMI in their system. However, for applications, it is desirable to have DMI in systems that are easily compatible with standard Si based technology. Moreover, molecular beam epitaxy is not an industry compatible growth method due to the lower growth rate. Hence, it is desirable to have DMI in FM layer grown by sputtering, which is a simple, yet versatile deposition method used in industry.

In this Rapid Communication, we demonstrate direct observation of iDMI in graphene/$Ni_{80}Fe_{20}$/Ta heterostructures grown on Si substrates by magnetron sputtering. We use asymmetric spin-wave dispersion probed by Brillouin light scattering, a technique which has already been established as a direct and reliable method of measuring the iDMI [14, 15, 19, 20]. We present a systematic study of iDMI as a function of the thickness of $Ni_{80}Fe_{20}$ and show that the iDMI in this system originate from the interface between $Ni_{80}Fe_{20}$ and graphene. Furthermore, by controlling the defects at the interface by Ar deposition pressure during growth of $Ni_{80}Fe_{20}$, we establish that the DMI in this system arises from the defect induced spin-orbit coupling. This is further supported by a correlation between the DMI and spin-mixing conductance, both of which are related to spin-orbit coupling.

We have used high-quality commercial CVD graphene (from Graphenea) on a Si/$SiO_2$ substrate. A series of samples consisting of substrate/graphene/$Ni_{80}Fe_{20}$ ($t$)/Ta (2), with $t$ = 3, 4, 6, 8, 10, 15 nm (digits indicate thickness in nm) were deposited at room temperature. NiFe/Ta bilayers were deposited with a varying thickness of NiFe using DC magnetron sputtering at varying Ar working pressure at 3 µTorr base pressure. The Ar working pressure for deposition of NiFe thin films on graphene is varied from 2 mTorr to 10 mTorr for intentionally tailoring the defects in the graphene layer. A set of reference samples of $Ni_{80}Fe_{20}$ ($t$)/Ta (2) were also simultaneously prepared on Si/$SiO_2$ substrates. The growth rate of NiFe is kept low at <1.4 Å/sec. The thickness of NiFe film was determined using x-ray reflectivity (XRR) technique for different Ar working pressure. The sputtering target was placed at an angle of 45° *w.r.t.* the surface of the substrate to minimize the possible bombardment of ions or neutral atoms on the graphene layer. The distance between the substrate and the sputtering target was approximately 8 cm.

In order to investigate the asymmetric spin-wave dispersion caused by iDMI, BLS measurements were performed in Damon-Eshbach (DE) geometry, *i.e.* by applying the magnetic field perpendicular to the plane of incidence of the laser beam. This allows for probing the spin waves propagating along the in-plane direction perpendicular to the applied field, *i.e.* in the DE geometry where the iDMI effect on the non-reciprocity in spin-wave frequency is maximal at the room temperature. The details of the BLS measurement can be found elsewhere [14, 19, 20, 50]. To get well defined BLS spectra for the larger incidence angles, the spectra were obtained after counting photons for several hours. We have also performed FMR measurements for calculating effective damping parameter ($\alpha_{eff}$) and spin-mixing conductance ($g_{\uparrow\downarrow}$) using a co-planer waveguide (CPW) based broadband FMR set-up for excitation frequencies of 2−12 GHz.

The magnetization hysteresis loop measured by vibrating sample magnetometer (VSM) at room temperature for the film stacks NiFe (10 nm)/Ta (2nm) and Gr/NiFe (10 nm)/Ta (2nm) are shown in Figure 1(a) and 1(b), respectively. It is observed that the value of saturation magnetization ($M_s$) in sample with graphene is little lower while the coercivity is a bit higher than the reference sample (without graphene underlayer). Fig. 1(c) shows the variation of effective magnetization ($M_{eff}$) (extracted from field dependent BLS measurement with the inverse of the NiFe thickness. A linear variation of $M_{eff}$ with inverse of NiFe thickness with negative slope is observed. The *y*-inercept refers to the value of saturation magnetization ($M_s$). The $M_s$ values obtained from field dependent BLS measurement agree well with that of VSM as well as FMR measurement. The Raman spectra of CVD grown graphene on a Si/SiO$_2$ before and after the deposition of NiFe (10 nm)/Ta (2 nm) bilayer thin films are shown in Figure **1(d)**. G and 2D peaks related to graphene are observed at 1590 cm$^{-1}$ and 2695 cm$^{-1}$, respectively. The D peak observed in Fig. **1(d)** (top panel) at 1349 cm$^{-1}$ corresponds to the defects induced due to sputter deposition of NiFe/Ta bilayer, which is well known and appears due to intervalley scattering [47]. The D' peak observed in Fig. **1(d)** (bottom panel) centered at 1620 cm$^{-1}$ is also a disordered induced Raman peak, which comes from intravalley scattering. Spectral weights of graphene modes were obtained by fitting to Lorentzian function. We use the spectral weight ratio ($I_D/I_G$ ratio) as a quantitative measure of sputtering damage. The $I_D/I_G$ ratio for graphene after deposition of FM/capping layer is 2.01 ± 0.02, which indicates that some defects are introduced in the graphene layer after deposition. We calculate the average crystallite size ($L_a$) using the relation: $L_a = (2.4 \times 10^{-10}) \lambda_{laser}^4 (nm) \left(\frac{I_D}{I_G}\right)^{-1}$, where $\lambda_{laser}$ = 514 nm is excitation wavelength of laser used for Raman measurement [47], average nanocrystallite size ($L_a$) of 8.33 nm is obtained in our case. The vibrating sample magnetometer (VSM) measurement was performed to estimate the values of the saturation magnetization ($M_s$) of the samples. All the films are in-plane magnetized as revealed by VSM measurement.

To quantify the strength of iDMI in these graphene-based heterostructures, we measured the spin-wave dispersion (*f* vs. *k*) relation at the applied in-plane field of *H* = 1 kOe. Here, transferred wave vector *k* was selected by changing the angle of incidence of the laser beam. The spin waves propagating in the opposite directions were simultaneously detected as Stokes and anti-Stokes peaks in the BLS spectra. In order to model our experimental data, we use the equation for the difference in the frequencies of counter-propagating spin waves given below [18]

$$\Delta f = \left\{ \frac{[f(-k, M_z) - (f(k, M_z))] - [f(-k, -M_z) - (f(k, -M_z))]}{2} \right\}; \quad (1)$$

$$\Delta f = \frac{2\gamma}{\pi M_s} Dk + \Delta\epsilon, \quad (2)$$

where $M_z$ denotes the magnetization along applied magnetic field. $\Delta\epsilon(k) = \epsilon(K_\perp, k) - \epsilon(K_\perp, -k)$, where $K_\perp$ is the interfacial magnetic anisotropy, describes a correction in frequency due to interface anisotropy and any other offset. The second term $\Delta\epsilon$ is very small compared to the first (DMI) term on the right hand side. The first term is linear in $D$ and $k$, where $D$ is the iDMI constant. Here $\gamma$ is the gyromagnetic ratio whereas $M_s$ is the saturation magnetization. Therefore, Eq. **(2)** provides a direct way to quantify the strength of iDMI by experimentally measured quantities $\Delta f$, $M_s$ and $k$.

Figure **2(a)** shows the schematic of the BLS measurement. As Eq. (2) suggests that the amount of frequency asymmetry is linear in $k$, typical BLS spectra for DE spin waves recorded at higher wave vector ($k$ = 18.1 rad/µm) under oppositely oriented external magnetic fields in graphene/$t$ Ni$_{80}$Fe$_{20}$/2 Ta, where $t$ = 3, 4, 6, 8 nm, are presented in Fig. **2(b)**. For clarity, anti-Stokes side of BLS spectra for oppositely oriented magnetic fields are shown with the corresponding Lorentzian fits. By looking at these BLS spectra, it is evident that the frequency difference between counter propagating spin waves ($\Delta f$), which is the measure of the strength of iDMI, decreases with increasing $t$ (NiFe thickness). In the case of $t$ = 3 nm, $\Delta f$ ≈ 0.24 GHz which is reasonably large for graphene/NiFe/Ta system. In addition, typical BLS spectra for reference sample (NiFe (3 nm)/Ta (2 nm)) is also presented in the lower panel of Fig. 2(b). Interestingly, almost negligible $\Delta f$ is observed for the reference sample which ruled out the possibility of DMI contribution from the interface of NiFe and Ta capping layer. It should be noted here that such a large non-reciprocity in spin-wave frequency is indicative of the presence of iDMI in these systems. Earlier, Hehn *et al.* have shown that frequency non-reciprocity in Pt/NiFe originated due to normal uniaxial interface anisotropy (NUIA) is weaker by an order of magnitude, which is almost negligible [16].

To quantify the magnitude of iDMI for various thicknesses of NiFe in graphene based heterostructures, full *k*-dependent BLS measurements have been performed by varying the angle of incidence of the laser beam under oppositely oriented magnetic fields. Figure **3(a)** represents the variation of $\Delta f$ as a function of the spin-wave wave vector ($k$) for samples with different $t$. The best linear fit to the data using Eq. **(2)** provides the quantitative estimation of iDMI. The linear correlation of $\Delta f$ with $k$ thus yields a direct way of measuring the magnitude of iDMI. A maximum $D$ of 67 ± 10 μJ/m$^2$ was observed for $t$ = 3 nm from the slope of the linear correlation. This observation of significant and sizeable $D$ in these graphene-based heterostructures, which we will attribute to the extrinsic spin-orbit coupling arising due to the presence of defects in graphene after FM/Ta deposition. In our case, conduction electron mediated Rashba-DMI mechanism may play a minor role as described recently by Yang *et al.* for graphene/FM systems as the thickness of the NiFe layers used in our experiment is much above the monolayer of NiFe [48]. Figure **3(b)** shows the variation of $D$, extracted from the procedure mentioned above as a function of the inverse of Ni$_{80}$Fe$_{20}$ thickness. Interestingly, an almost linear scaling behaviour of $D$ with the inverse of ferromagnetic thickness is observed, which indicates its origin primarily from the graphene/FM interface. To further confirm the presence of iDMI in these heterostructures, reference samples (without graphene underlayer) were also measured. Importantly, almost negligible frequency asymmetry is observed in the reference sample..

Next, we investigate the effect of defects induced in graphene layer achieved by varying Ar pressure during NiFe layer deposition. Shown in Figure **4(a)** are the Raman spectra taken after 10 nm of NiFe deposition at various Ar pressure (Raman spectra for 3 nm of NiFe are given in the Supplementary Material, Fig. S2). There is a substantial change in the Raman $I_D/I_G$ ratio with Ar pressure as evident by analyzing the Raman data. We notice that the $I_D/I_G$ ratio increases from 201% to 227% when Ar pressure increases from 2 mTorr to 10 mTorr. The $I_D/I_G$ ratio and $L_a$ for graphene/NiFe bilayer thin films with varying Ar pressure are summarized in Table I. This observed behaviour can be explained by taking into account of the impact of Ar neutral atoms similar to Chen *et al.*, where a similar increase of $I_D/I_G$ ratio was found with Ar pressure [47]. The increase of defects with Ar pressure is also expected to give rise to a defect induced extrinsic spin-orbit coupling. In fact, theoretical calculations showed that $sp^3$ distortion induced by an impurity can lead to a large increase in the spin-orbit coupling in graphene [51]. There are also several experimental reports that show the presence of spin-orbit coupling in graphene by hydrogenation, fluorination or the presence of adatoms [52-54].

**TABLE I.** Various parameters as a function of Ar working pressure during deposition of NiFe on graphene. $I_D/I_G$ ratio is obtained from Raman spectra. Average crystallite size ($L_a$) is estimated from Raman spectra. Saturation magnetization ($M_s$) estimated from field dependent BLS measurement. Spin-mixing conductance ($g_{\uparrow\downarrow}$) is obtained from FMR measurement. The strength of surface DMI ($D_s$) is determined from BLS technique.

| Ar working pressure (mTorr) | $I_D/I_G$ | $L_a$ (nm) | Saturation Magnetization (emu/cc) | Spin-mixing conductance (m$^{-2}$) × 10$^{19}$ | Surface DMI constant (J/m) × 10$^{-15}$ |
|---|---|---|---|---|---|
| 2 | (2.01 ± 0.02) | 8.33 ± 0.08 | 708 | 0.12 ± 0.05 | 93 ± 10 |
| 6 | (2.04 ± 0.02) | 8.21 ± 0.07 | 670 | 0.47 ± 0.10 | 143 ± 10 |
| 10 | (2.27 ± 0.02) | 7.38 ± 0.06 | 606 | 1.18 ± 0.10 | 193 ± 10 |

To investigate the effect of the defects on iDMI for graphene/NiFe based heterostructures, the variation of $\Delta f$ as a function of spin-wave wave vector ($k$) for samples with $t$ = 3 nm deposited at different Ar pressure is shown in Fig. **4(b)**. The data is well fitted with Eq. **(2)**. The $M_s$ values used in the fitting are shown in Table 1. These values are determined using BLS and FMR measurements and cross checked using vibrating sample magnetometer. The differennce in slopes of the linear fittings to the data points results from the variation of *D* in these samples. This variation of iDMI is likely due to the defect induced extrinsic spin-orbit coupling of graphene.

To further understand this behavior, we determined spin pumping from NiFe layer to graphene by FMR measurements. Similar to iDMI, spin pumping is also an interfacial effect, and hence a correlation between the two is expected, which is recently established in ferromagnet/heavy metal systems [18]. According to the theory of spin pumping, the effective damping parameter, $\alpha_{eff}$ of the FM is given by:

$$\alpha_{eff} = \alpha_0 + \alpha_{SP} \tag{3}$$

here $\alpha_0$ is the intrinsic Gilbert damping, whereas $\alpha_{SP}$ is the damping due to spin-pumping effect, which is given by [2]:

$$\alpha_{SP} = \frac{\gamma h g_{\uparrow\downarrow}}{4\pi M_s t_{NiFe}}, \tag{4}$$

where, $g_{\uparrow\downarrow}$ is the spin-mixing conductance, and *h* is the Planck's constant. Other parameters carry the same meaning as discussed earlier in this paper.

From the FMR data we determine, the spin-mixing conductance ($g_{\uparrow\downarrow}$) by performing NiFe thickness dependent study. In comparison to our reference samples (see Supplementary Material, Fig. S1), an enhancement of damping parameter was observed in the sample with graphene consistent with Eq. **(3)**. This enhancement can be attributed to spin pumping as previously reported for this system [55]. Figure **4(c)** shows the variation of $\alpha_{eff}$ with the inverse of NiFe thickness ($1/t_{NiFe}$) for the samples deposited at various Ar pressure. From the linear fitting of the effective damping parameter versus the inverse of NiFe thickness with Eq. **(3)**, we determine $g_{\uparrow\downarrow}$. The obtained values of $g_{\uparrow\downarrow}$ with different Ar pressure for the graphene/NiFe samples are displayed in the Table I. We see a clear increase of $g_{\uparrow\downarrow}$ with Ar pressure indicating an enhanced spin pumping with Ar pressure. This enhanced spin pumping with Ar pressure can be attributed to defect induced spin-orbit coupling at the interface. It is observed that surface DMI constant $D_s$ ($D*t_{NiFe}$) and $g_{\uparrow\downarrow}$ both increase with the increase in Ar pressure, thus indicating a direct correlation between these two quantities. Figure **4(d)**, shows a plot of $D_s$ with spin-mixing conductance $g_{\uparrow\downarrow}$, which clearly shows this correlation. This correlation, which is already established in the FM/HM system [18] and antiferromagnet (AFM)/FM system [25] further supports our claim that defect induced extrinsic spin-orbit coupling at the interface is the primary origin of iDMI in our system.

In summary, we have systematically studied the FM layer thickness dependence of iDMI in technologically important graphene/$Ni_{80}Fe_{20}$/Ta heterostructures using BLS spectroscopy. By measuring the spin-wave frequency non-reciprocity, we observe a sizeable iDMI in these stacks. FM thickness variation reveals that $D$ varies linearly with the inverse of NiFe thickness demonstrating its purely interfacial origin. Furthermore, we observe a tunability of surface DMI constant and spin-mixing conductance with defect density in the graphene layer obtained by varying Ar pressure during DC sputtering of NiFe. Remarkably, there is a direct correlation between surface DMI constant and spin-mixing conductance, while both are found to be correlated with the increase in defect density. Hence, we conclude that defect induced extrinsic spin-orbit coupling may play a major role in the observed iDMI in these samples. Our detailed FM layer thickness and Ar pressure dependent study of iDMI will enrich the understanding on the observation and tunability of iDMI in these 2D heterostructures for controlling chiral spin structure and magnetic domain-wall based magnetic storage, memory and logic devices.

We acknowledge Dr. Saroj Dash for useful discussions. A.B. gratefully acknowledge the financial assistance from Department of Science and Technology (DST), Government of India under grant no. SR/NM/NS-09/2011 and S. N. Bose National Centre for Basic Sciences under project no. SNB/AB/18-19/211. P.K.M. acknowledges partial support from the Ministry of Human Resource Development under the IMPRINT program (Grants No. 7519 and No. 7058), the Department of Electronics and Information Technology (DeitY), and the Department of Science and Technology under the Nanomission program. A.K.C. acknowledges DST, Government of India for INSPIRE fellowship (IF150922), while A.K. acknowledges support from the Council of Scientific and Industrial Research (CSIR), India.

# Figures

**Figure 1:**

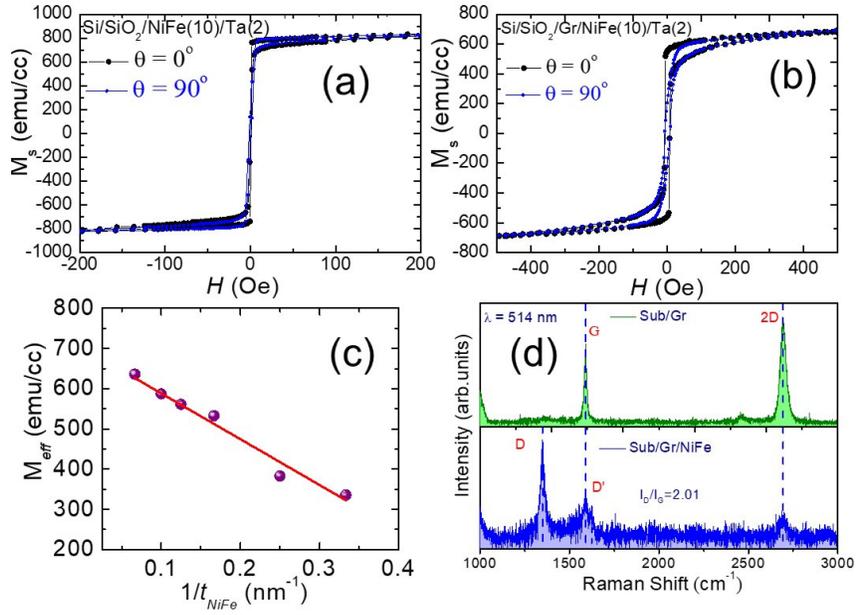

**Figure 2:**

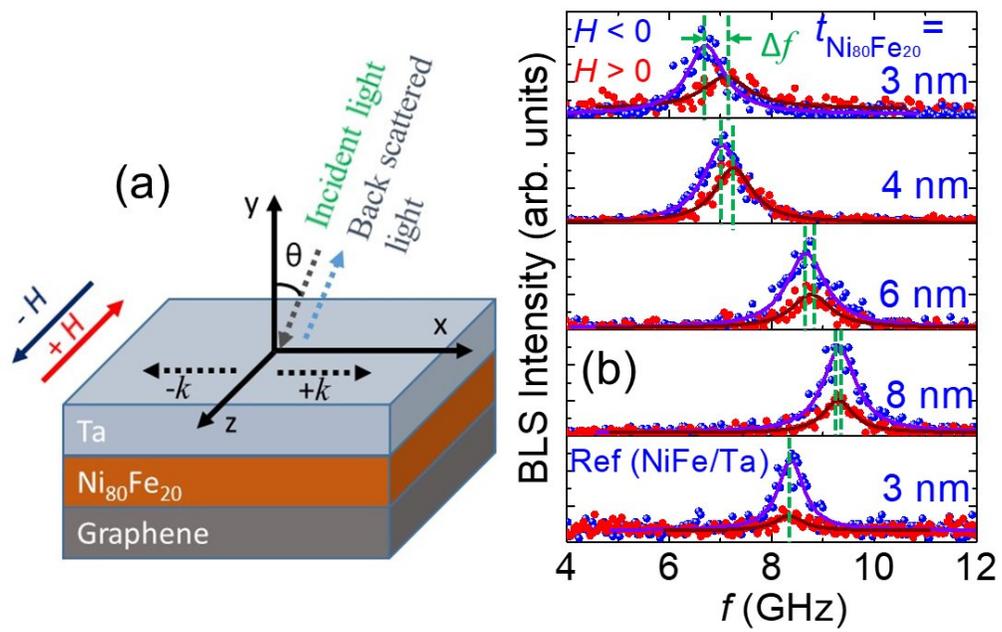

**Figure 3:**

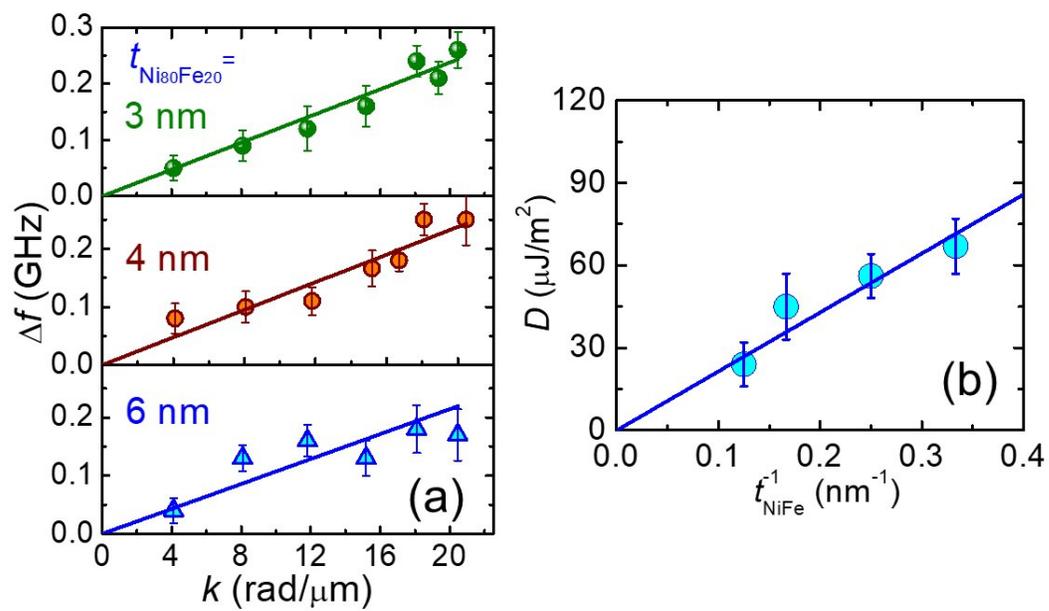

**Figure 4:**

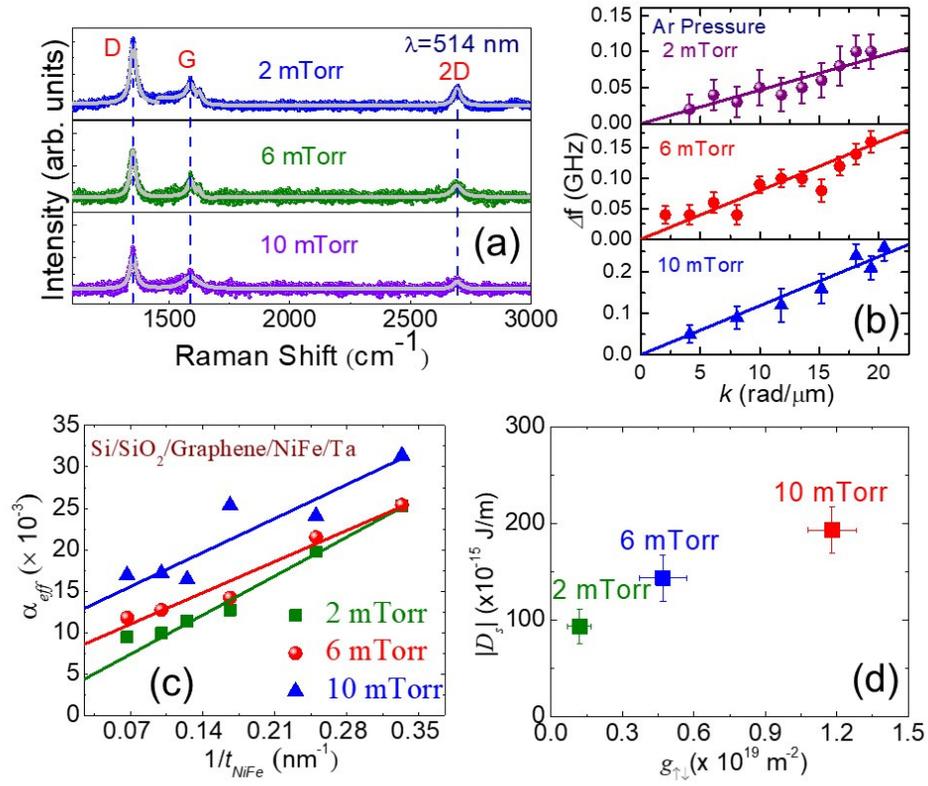

**Figure Caption:**

**Figure 1:** Magnetization hysteresis loop for sample stack (a) NiFe (10 nm)/Ta (2 nm) (b) Gr/NiFe (10 nm)/Ta (2nm) (deposited at 2 mTorr Ar working pressure) with magnetic field applied within the film plane. Here θ refers to the angle between two mutually perpendicular direction within the sample plane. (c) Variation of $M_{eff}$ (extracted from magnetic field dependence using BLS) as a function of inverse of NiFe thickness in Gr/NiFe/Ta system. Red solid line is the best linear fit. (d) Raman spectra of CVD grown graphene on a Si/SiO$_2$ before (top panel) and after (bottom panel) the deposition of NiFe (10 nm)/Ta (2 nm) bilayer thin films.

**Figure 2:** (a) Schematic of the film stack along with BLS geometry. (b) Representative BLS spectra for DE spin waves acquired at a fixed wave vector $k$ = 18.1 rad/μm under oppositely oriented external applied fields $H$ = 1 kOe in graphene/Ni$_{80}$Fe$_{20}$ ($t$)/Ta (2 nm) sample (deposited at 10 mTorr Ar pressure) (top four panels) and reference sample NiFe (3 nm)/Ta (2 nm) (at the bottom most panel) for two counter propagating directions of spin waves. The spectrum corresponding to particular thickness of Ni$_{80}$Fe$_{20}$ is indicated by mentioning thickness value in each panel. Solid curve is the fit using Lorentzian function.

**Figure 3:** (a) Plot of $\Delta f$ vs $k$ for graphene/Ni$_{80}$Fe$_{20}$ ($t_{NiFe}$)/Ta (2 nm) samples with various values of $t_{NiFe}$. Symbols represent the experimental data points and solid lines are the fit using Eq. **(2)**. (b) Variation of $D$ with the inverse of Ni$_{80}$Fe$_{20}$ thickness. The error bar in $\Delta f$ is shown by considering the error from the fitting of spectra as well as the instrumental resolution to determine the peak frequency and for $D$, errors in $M_s$ as well as $\Delta f$ have been taken into account. The red solid line is the linear fit.

**Figure 4:** (a) Raman spectra of CVD grown graphene on a Si/SiO$_2$ after the deposition of NiFe (10 nm)/Ta(2 nm) bilayer thin films at different Ar pressure. (b) Plot of $\Delta f$ vs $k$ for graphene/NiFe (3 nm)/Ta (2 nm) samples deposited at various Ar pressure. Symbols

represent the experimental data points and solid lines are the fit using Eq. (2). (c) Variation of effective damping constant with the inverse of NiFe thickness for various Ar pressure investigated from FMR measurement. (d) Plot of surface DMI constant with spin mixing conductance. Ar pressure value is mentioned on each data point.